# The response of an electron in the image potential on the external steady fields

P.A. Golovinski [1,2], M.A. Preobrazhenskiy [2]

[1] *Moscow Institute of Physics and Technology (State University), 141700, Moscow, Russia*

[2] *Voronezh State University of Architecture and Civil Engineering, 394006, Voronezh, Russia*

*e-mail: golovinski@bk.ru*

The exact wave functions, which describe the states of an electron, bound in the image potential, and the magnetic field, which is perpendicular to surface of a metal, are obtained. The correction terms to the energy of an electron in the first order of the perturbation theory with respect to the interaction with the external electric field are calculated. The possibility of experimental measurement of a local electric field by the level shift is analyzed.

Low-dimensional systems (two-dimensional, one-dimensional and zero-dimensional) have a number of unusual properties that largely determine their role in the study and developing of nanoscale devices [1]. In particular, an electron near the metal surface interacts with induced on the surface polarization charge, attraction to which leads to appearance of localized states. Far from the surface in a vacuum this potential is well approximated by the classical long-range potential [2]. An electron near metal surface interacts with the field of its mirror image [3-5 interaction with which is asymptotically described by a one-dimensional Coulomb operator of an electron energy for the attraction to the metal surface [6] (the atomic system of units is used: $\hbar = e = m = 1$). The model is corresponded to the "one-dimensional atom" with an effective charge $Z = 1/4$, and the hydrogen-like electron states with energies $E_\nu = -1/32(\nu + a)^2$ are determined by the quantum-number $\nu$ and the quantum defect $a$, which weakly depends on $\nu$ v [3, 7]. This theoretical model [8-10] is confirmed by spectroscopic observations, such as two-photon resonance emission of electrons from the metal [11]. More sophisticated accounting of electron interaction with the surface plasmons is possible within the framework of multi-particle theory and is important for correct description of the width of the lower states with $\nu \leq 3$. In this paper the response of an electron that is bound in the image potential on electric and magnetic fields is considered and the possibility of implementation this effect in spectroscopic measurement of electric field near the surface of metal is analyzed.

Nonrelativistic Hamiltonian of an electron in electromagnetic field can be expressed in terms of the operators of the momentum, the intrinsic magnetic moment, vector **A** and scalar $\varphi$ potentials of electric field and magnetic field strength **H** [12] by equation:

$$\hat{H} = \frac{1}{2}\left(\hat{\mathbf{p}} + \frac{\mathbf{A}}{c}\right)^2 - \hat{\boldsymbol{\mu}}\mathbf{H} - \varphi. \tag{1}$$

Let's consider behavior of an electron in the image potential in homogeneous electric and magnetic fields, which are directed perpendicular to a metal surface.

In this case, the scalar potential depends only on the coordinate $z$ and it is determined by external electric field strength by equation $\varphi = 1/4z + Fz$. The vector potential can be taken in the form $\mathbf{A} = \{-Hy, 0, 0\}$. Then, stationary Schrödinger equation has the form

$$\{(\hat{p}_x^2 + \hat{p}_y^2 + \hat{p}_z^2 + 2\hat{p}_x Hy/c + H^2 y^2/c^2)/2 - Fz - 1/4z - \sigma H/c - E\}\psi(x, y, z) = 0. \tag{2}$$

It is assumed here that the operator $\hat{\boldsymbol{\mu}}\mathbf{H}$ commutes with the Hamiltonian (1) and, therefore, the electron has a definite value of the spin projection. Hamiltonian of the Eq. (2) is additive with respect to the Cartesian coordinates. Hence, variables in the solution are entirely separated.

Furthermore, because the Hamiltonian of the Eq. (2) does not contain the variable $x$, it commutes with the component of the momentum operator $\hat{p}_x$. The corresponding component of the wave function is equal to the eigenfunction of operator $\hat{p}_x$ normalizaed on $\delta$-function it is described by a plane wave $\exp(ip_x x)/\sqrt{2\pi}$. With this in mind, wave function of electron can be written as

$$\psi(x, y, z) = \exp(ip_x x)\,\chi(y)\,\varphi_\nu(z)/\sqrt{2\pi}. \tag{3}$$

Since the variables in the wave function (3) are divided, the total energy of the electron has the form:

$$E = p_x^2/2 + E_y + E_z. \tag{4}$$

We obtain two independent equations for $y$ and $z$ dependent components of the wave function by substituting the wave function (3) into Eq. (2):

$$\{d^2/dy^2 - H^2(y + cp_x/H)^2/c^2 + p_x^2 + 2E_y + 2\sigma H/c\}\chi(y) = 0, \tag{5}$$

$$\{d^2/dz^2 + 2(Fz + 1/4z + E_z)\}\varphi_z(z) = 0. \tag{6}$$

Introducing new variable $Y = y + cp_x/H$ and parameters

$$E_m = p_x^2/2 + E_y + \sigma H/c; \quad \omega = H/c \tag{7}$$

allows us to reduce the equation (5) to the Schrödinger equation for a harmonic oscillator [12]

$$(d^2/dY^2 + 2E_m - \omega^2 Y^2)\chi(Y) = 0. \tag{8}$$

Component of the wave function $\chi(Y)$ is expressed in terms of Hermite polynomials $H_m$ [12, 13] in the form:

$$\chi(Y) = (H/\pi c)^{1/4} (2^n m!)^{-1/2} \exp(-HY^2/2c) H_m(Y\sqrt{H/c}). \tag{9}$$

Because equidistant discrete spectrum is

$$E_m = (m + 1/2)\omega; \quad m = 1,2,3,\ldots,\infty, \tag{10}$$

and following Eq. (7) and Eq. (9), we obtain the energy component $E_y$ that is linear on the magnetic field strength:

$$E_y = (m - \sigma + 1/2)H/c - p_x^2/2. \tag{11}$$

Thus, in the image potential field and magnetic field, which is perpendicular to the surface of metal, one has linear Zeeman.

The value $E_z$ is determined by the Eq. (6) in the first-order of the perturbation theory with respect the interaction of an electron with an external electrostatic field. In the absence of an external electric field function $\varphi_\nu(z)$ satisfies the equation

$$(d^2/dz^2 + 1/2z + 2E_\nu^0)\varphi_\nu(z) = 0. \tag{12}$$

Solution of the Eq. (12) is equal to $\varphi_\nu(z) = f_\nu(z)/z$, where $f_\nu(z)$ coincides with the radial part of the wave function of $s$ state of a hydrogen-like atom with charge $Z = 1/4$ [12]. For negative values of energy $E_\nu < 0$ solutions of the Eq. (12), normalized to unity, are expressed in terms of confluent hypergeometric function $\Phi(1-\nu, 2, 2z/\nu)$ [13]

$$\varphi_\nu(z) = 2\nu^{-3/2} z \exp(-z/\nu) \Phi(1-\nu, 2, 2z/\nu). \tag{13}$$

Quantum number $\nu = 1/\sqrt{-2E_\nu^0}$ is positive integer number.

The dipole moment of an electron $z_\nu = \langle \varphi_\nu(z) | z | \varphi_\nu(z) \rangle$ in the image potential is nonzero because it has no mirror symmetry. Therefore, in contrast to the three-dimensional Coulomb potential, the level shift is defined by the expression

$$\Delta E_z = F \cdot z_\nu \tag{14}$$

in the first-order of the perturbation theory with respect to the interaction of an electron with an external electric field [14].

The calculation of the cross sections for single-photon transitions in the Coulomb potential between states with different energies $E_i$ and $E_f$ is usually based on the integral equations for the confluent hypergeometric functions [15], and the dipole moment of a state is obtained by a limit $E_i \to E_f$. For the static fields a direct algorithm for calculating the dipole moment of a discrete state, based on the reduction of expression for the matrix element $z_\nu$ in Eq. (14) to the superposition of the Euler integrals [13], can be formulated. With taking into account the normalization condition (13), it turns to the form

$$z_v = \frac{3}{2}\left\{1+(v-1)!\sum_{s=0}^{\min[1,v-2]} \frac{(-3-s)(-2-s)...(-2+s)}{(v-2-s)![(s+1)!]^2(s+2)!}\right\}. \quad (15)$$

The sum in the Eq. (15) for the states with quantum number $v > 2$ contains two terms. Calculation leads to the relation

$$z_v = 3v^2/2, \quad (16)$$

that formally coincides with the average value of a distance $r$ for the $s$-state in three-dimensional hydrogen atom. It is also true for the ground and first excited state.

Taking into account relation between quantum number $v$ and unperturbed energy $E_v^0$, and Eq. (16), $z$-component of the bound-state energy in an electrostatic field can be represented in the form

$$E_z = -1/2v^2 + 3Fv^2/2. \quad (17)$$

The condition of validity for the first-order perturbation theory is determined by the relative smallness of the energy shift of the neighboring states compared with the distance between them, which leads to the condition $F \ll (2v+1)/(3v^4(v+1)^2)$. For levels with $v \gg 1$ it simplifies and takes a form $F \ll v^{-5}$. Narrowing the range of applicability of the perturbation theory near the boundary of ionization is explained by the increased length of the electron cloud. Substituting $y$-component and the $z$-component of the energy from Eq. (11) and Eq. (17) into Eq. (4) for the total energy of the electron, one can obtain equation

$$E(m,v) = (m - \sigma + 1/2)H/c - 1/2v^2 + 3Fv^2/2. \quad (18)$$

Since the energy (18) does not depend on the component $p_x$ of the momentum, the electron states, which are described by the wave functions (3), have an infinite degeneracy typical for the Landau levels.

Dependences of energy shift (18) on the electric field strength for a number of highly excited states of an electron in the image potential is shown in Fig. 1. These dependences demonstrate a linear rise of the energy distances between the levels with increasing the electric field strength.

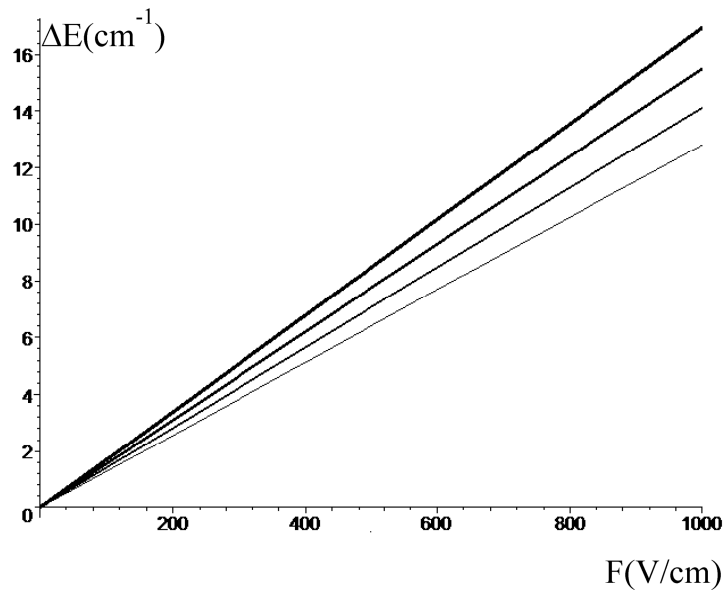

Fig.1. The dependences of the energy shift $\Delta E$ of electron states, bound in the image potential, for quantum numbers (bottom to top) $v = 20, 21, 22, 23$, on the electric field strength.

As it follows from Eq. (18), linear static Stark effect contributes to the energy of all states, which are bound by one-dimensional Coulomb interaction. In contrast, in three-dimensional Coulomb potential, the ground state is not shifted in a uniform electric field. This difference is determined by difference in the physical cause of the energy shift – in three-dimensional potential shift, which is linear on the field, it is caused by interaction of degenerate levels, and in the one-dimensional case it is caused by the mirror asymmetry of states.

In accordance to the Eq. (18), the change in energy of transition between states with the equal quantum numbers $m$ is determined by the electric field strength and quantum numbers $v_i$ and $v_f$ of levels, between which the radiation transition happens:

$$\Delta E = 3F\left(v_i^2 - v_f^2\right)/2 \ . \qquad (20)$$

The error in determining the electric field strength by spectroscopic measurements of the Rydberg states with $v=90$ with high resolution in ultracold gas of cesium atoms is equal to 1 mV/cm [16]. The resolution is determined by a width of the line of the high-frequency resonance, which is equal to $\delta E = 20 \div 60$ kHz ($3 \cdot 10^{-6}$ cm$^{-1}$).

Changing the energy of transition (in units cm-1) between the Rydberg states in the image potential, when quantum numbers differ by $k$, is equal to $\Delta E_k \approx 6.42 \cdot 10^{-5} v \cdot k \cdot F$ (the electric field strength is expressed in V/cm and it is assumed, that $k << \nu$ )). The minimum electric field strength, measured by the spectroscopic methods in the image potential, limited by condition $\Delta E_k \geq \delta E$, is equal to $F_{\min} \geq 5 \cdot 10^{-2} / \nu k$. Thus, in the image potential field, the resolution of 1 mV/cm can be achieved at the transition between adjacent levels with quantum numbers $\nu \sim 50$ and can be increased by increasing photon energy.

In addition, the spectroscopic measurements in the image potential field are simplified (compared to the case of the Rydberg states of atoms) due to a more simple energetic structure of the bound states without quasidegenerate levels. In contrast to the electric field, the accuracy of the spectroscopic measurements of the magnetic field does not increase with increasing of the quantum numbers of the discrete states. However, the presence of the magnetic field facilitates the confinement of electrons, which allows localize the states in some space near the metal surface.

This research has been supported by the RFFI through Grant № 13-07-00270 and Ministry of Education and Science of Russian Federation state order № 2014/19-2881.